\documentclass[aps, superscriptaddress, twocolumn, nofootinbib]{revtex4}
\usepackage{amsmath, amssymb, mathtools, esint}
\usepackage[utf8]{inputenc} 
\usepackage{graphicx, fancybox}
\usepackage{epstopdf}
\usepackage{color}
\usepackage{cancel}
\usepackage{listings}
\usepackage{rotating}
\usepackage{hyperref}
\usepackage{filecontents}

\newcommand{\vb}{\;\vec B}
\newcommand{\vm}{\;\vec \mu}

\begin{document}
\title{Nonlinear dynamic interpretation of quantum spin}
\author{J.J. Heiner}
\email[]{hjoshuaj@gmail.com}
\affiliation{Department of Physics and Astronomy, University of Wyoming, WY United States} 
\author{H.C. Shaw}
\email[]{harry.c.shaw@nasa.gov}
\affiliation{NASA Goddard Space Flight Center, Greenbelt, MD United States}
\author{D.R. Thayer}
\email[]{drthayer@uwyo.edu}
\affiliation{Department of Physics and Astronomy, University of Wyoming, WY United States}
\author{J.D. Bodyfelt}
\email[]{jdbodyfelt@gmail.com}
\affiliation{Centrality, 48 Emily Place, Auckland, New Zealand, 1010}
\begin{abstract}
In an effort to provide an alternative method to represent a quantum spin, a precise nonlinear dynamics semi-classical model is used to show that standard quantum spin analysis can be obtained.  The model includes a multi-body, anti-ferromagnetic ordering, highly coupled quantum spin and a semi-classical interpretation of the torque on a spin magnetic moment in the presence of a magnetic field.  The deterministic nonlinear differential coupling equation is used to introduce chaos, which is necessary to reproduce the correct statistical quantum results.
\end{abstract}
\maketitle
\section{Introduction}
In recent publications it was discussed that it may be possible to understand the quantum mechanical spin states in a similar method used in deterministic chaos \cite{Thayer_Jafari, Thayer}.  Recently, a 2D nonlinear semi-classical perturbation model of the spin interaction with a magnetic field was developed \cite{2D_qspin} and expanded to a 3D exact model \cite{New_Zealand}.  Although the latter model lacked the needed chaos and the ability to replicate quantum statistics, it led to the concept that it might be possible to exhibit chaos internally in the spin model in order to reproduce the quantum mechanical spin results.  It is thus suggested to treat the quantum spin as a multi-body quantum spin (with many sub-element components).

Scientists have questioned and tested if quantum mechanics is indeed random \cite{Rauch, Bierhorst, Raffaelli}, or if our current understanding is not complete \cite{Einstein, Caves, Jaynes}.  If a black box was placed over a roulette wheel, the outcome would appear random.  However, upon lifting the box the realization is the outcome appears to be random due to chaos and sensitivity to initial conditions \cite{Gallow}.  Furthermore, recent research shows and measures a finite time evolution of a quantum mechanical state into an excited state (which previously had been considered instantaneous) \cite{Ossiander}.  It is with this viewpoint, that a proposed highly coupled, multi-body, anti-ferromagnetic ordering, semi-classical quantum spin model represent a single quantum spin (the evolution of which happens in a finite time) and that outcomes due to measurement are emergent behavior of chaos \cite{Thayer_Jafari}.

\begin{figure}
\includegraphics[scale=0.35, trim={0 7cm 0 8cm}]{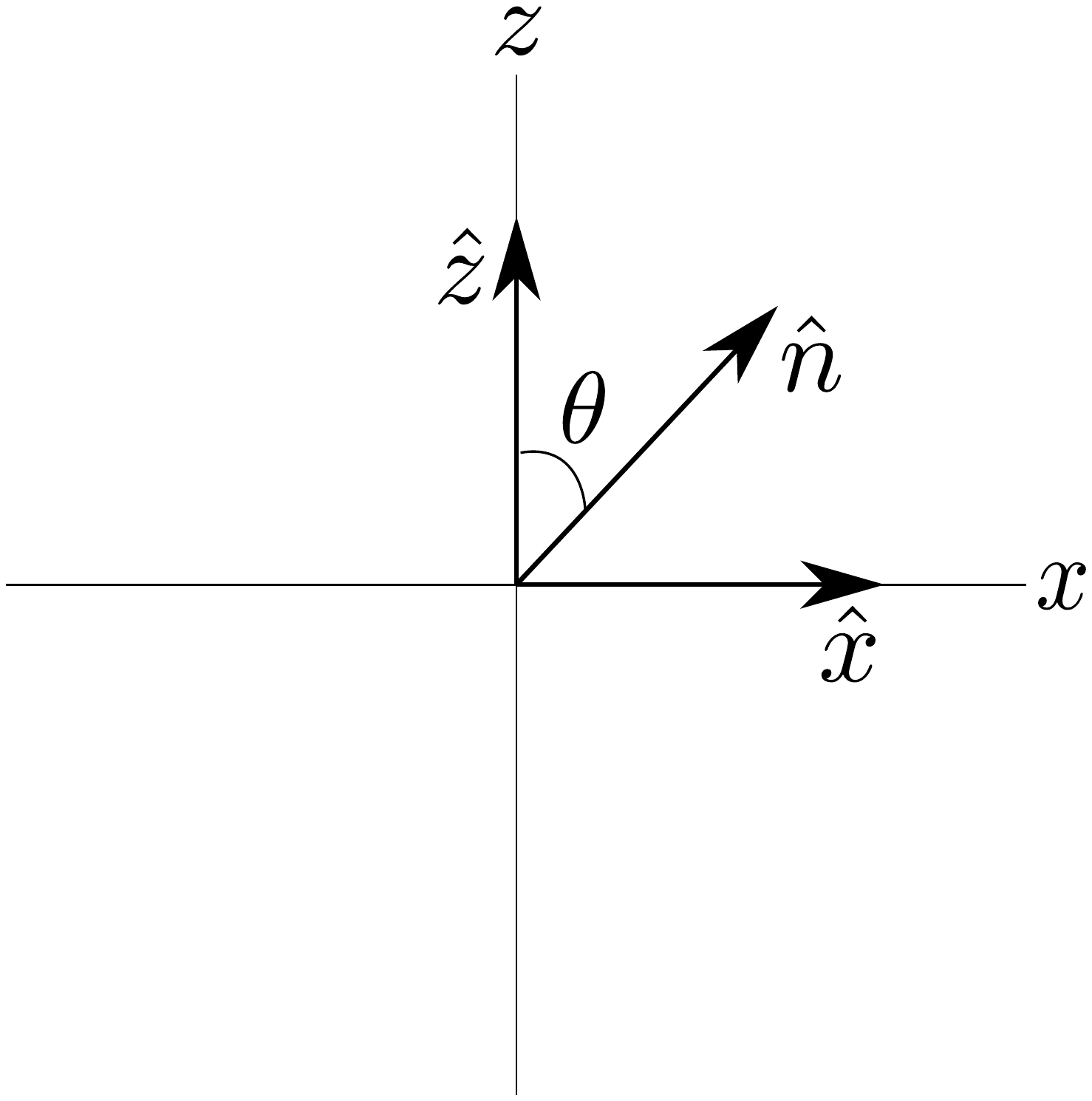}
\caption{\label{fig:2d_Geometry_Spin}2D Geometry for an individual component of the multi-body semi-classical spin model}
\end{figure}

The geometry used to describe the relationship between the quantum spin, $\mu$, and the magnetic field, $B$, can be seen in figure \ref{fig:2d_Geometry_Spin}, where the unit vector of the quantum spin is $\hat n$ and the magnetic field is constant, pointing in the $\hat z$ direction.

The probability that the quantum spin will collapse in the direction of the magnetic field, spin up, and the probability that it will collapse in the opposite direction, spin down, is given as

\begin{equation}
\begin{split}
P_\uparrow &= \cos^2 \theta/2, \\ P_\downarrow &= \sin^2 \theta/2.
\end{split}
\label{eq:Percentage}
\end{equation}

The probabilities shown in equation \ref{eq:Percentage} will be the testing criteria for the semi-classical quantum spin model.  In essence, the model will be given a finite time to evolve (the dynamics of which are chaotic) and settle into one of two states.
\section{Semi-Classical Quantum Spin Torque in a magnetic field}
As in past research, the semi-classical quantum spin is given a uniquely modified dual stable equilibrium torque model \cite{2D_qspin, New_Zealand}.  The classical torque model can be modified to exhibit two stable equilibria: $\theta = 0$ and $\theta = \pi$.  A function that fills this unique semi-classical torque behavior is,

\begin{equation}
\vec \tau_\textrm{sc} = \mu B \sin (2\theta) \, \, \hat m,
\label{eq:torque_semi_classical}
\end{equation}

\noindent where

\begin{equation}
\hat m = \frac{\hat n \times \vb}{\lvert \hat n \times \vb \rvert},
\end{equation}

\noindent and $\vm = \lvert \mu \rvert \hat n$.

This semi-classical torque function is similar to the one used in previous research \cite{2D_qspin, New_Zealand}, but the exact curve of the semi-classical torque had little affect on the results.  The most important feature is to have two stable equilibria locations as previously described.

Although equation \ref{eq:torque_semi_classical} implies the energy difference between the two equilibria states is zero, this is not an issue as a first order approximation justification is given in the conclusion section due to the multi-body coupling strength.

\section{Multi-Body Quantum Spin}

There are numerous theories on a maximal elementary particle mass \cite{Markov, Kadyshevsky, Mateev, Rodionov}; however, there is a lack of a limiting mass theory and in the Quantum Field Theory (QFT) mass can continously approach zero \cite{Hooft}.  Nevertheless, fundamental/elementry particle mass should be dissectable into an infinite number of pieces.  The quantum spin will be represented as many sub-element components that all together comprise the quantum spin.  This approach also resonates with the concept that the spin model should be considered as being infinitely complex \cite{Thayer_Jafari}.

\begin{figure}
\includegraphics[clip, trim={0cm 0cm 0cm 0cm}, scale=.65]{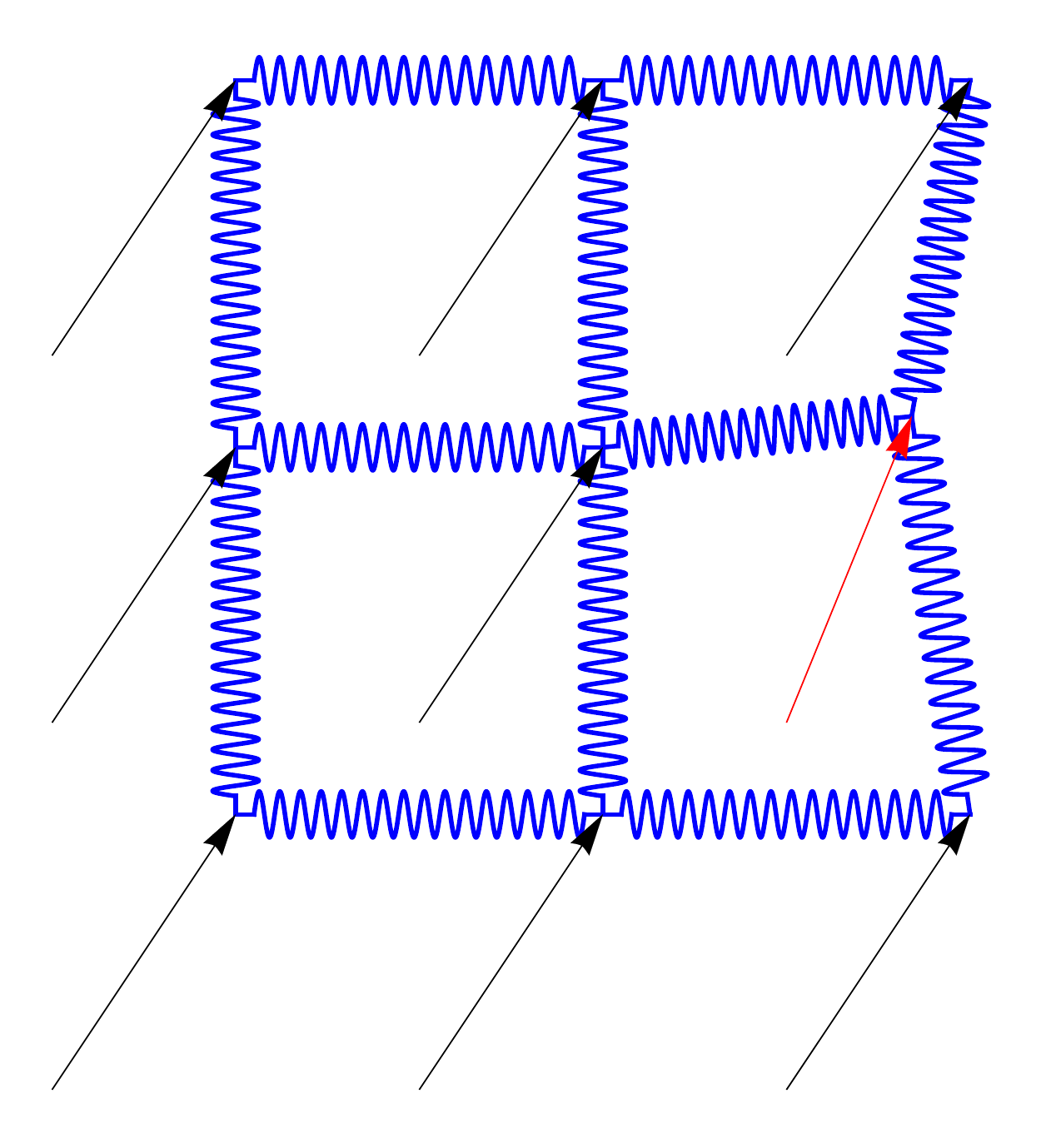}
\caption{\label{fig:Multi_Body_Model}Multi-body quantum spin model with ferromagnetic ordering if the blue coupling spring's equilibrium is the same as the distance between the sub-spin components.  In the model the distance between the individual spin components goes to zero to approximate a point particle.  A slightly perturbed spin component is demonstrated in red.}
\end{figure}

The multi-body quantum spin model with ferromagnetic ordering is shown in figure \ref{fig:Multi_Body_Model}.  For simplicity this model is shown; however, an anti-ferromagnetic ordering model is used in this research.  This can be obtained by allowing the coupling spring equilibrium to be double an individual quantum spin component length (assuming the distance between neighboring spin components goes to zero).

The torque on $\hat n$ due to nearest neighbor, $nn$, coupling (with the above assumption) is expressed as 

\begin{equation}
\vec \tau_\textrm{nn} =  \lvert \mu \rvert \hat n \times \vec F_{\hat n},
\label{eq:torque_nn}
\end{equation}

\noindent where

\begin{equation}
\vec F_{\hat n} = \sum_{i=1}^{nn} k(\lvert \hat n_i - \hat n \rvert - a) \frac{\hat n_i - \hat n}{\lvert \hat n_i - \hat n \rvert} , 
\label{eq:force_nn}
\end{equation}

\noindent $k$ is the spring coupling constant, and $a$ is the spring equilibrium length ($a=0$ for ferromagnetic ordering and $a=2\lvert \hat n \rvert = 2$ for anti-ferromagnetic ordering).

\section{Equations of Motion}

The equation of motion and resulting angular acceleration, $\vec \alpha_{\hat n} =\dot{\vec \omega}_{\hat n}$, for an individual quantum spin component can be represented as

\begin{equation}
I \vec \alpha_{\hat n} = \vec \tau_{\hat n}, 
\label{eq:EoM_torque}
\end{equation}

\noindent where the torque term is the sum of all torques on the indivdual spin.  Included in the torque terms is a linear angular dissipation force, i.e. $\vec \tau_\textrm{diss} = -b\, \vec \omega_{\hat n}$, so that

\begin{equation}
\vec \tau_{\hat n} = \vec \tau_\textrm{sc} + \vec \tau_\textrm{nn} + \vec \tau_\textrm{diss}, 
\label{eq:torque_simplified}
\end{equation}

\noindent and $b$ is a dissipation factor.

An indiviual quantum spin component's equation of motions are

\begin{equation}
\begin{split}
\dot{\vec\omega}_{\hat n} =& \, \, I^{-1}(\mu B \sin (2\theta) \, \, \hat m\\ &+ \lvert \mu \rvert \hat n \times \sum_{i=1}^{nn}( k(\lvert \hat n_i - \hat n \rvert - a) \frac{\hat n_i - \hat n}{\lvert \hat n_i - \hat n \rvert})\\ &+ b\, \vec \omega_{\hat n}),\\
\dot{\vec \theta}_{\hat n} =& \, \, \vec\omega_{\hat n}.
\end{split}
\label{eq:alpha_total}
\end{equation}

The dissipation force is held to zero for some time and then turned on to represent the need for the quantum spin to fully collapse into one of two states: either aligned with the magnetic field, up, or anti-aligned with the field, down.

\section{Code}
The Dormand-Prince Runge-Kutta method is used to iterate the equations of motion, equation \ref{eq:alpha_total}, for each individual quantum spin component.

Initial combined averaged spin orientation values, $\left\langle \theta_{int} \right\rangle$ are divided equally from 0 to $\pi$ in increments of $\pi/1000$.  Each individual $\theta_i$ value is given a random value based around $\left\langle \theta_{int} \right\rangle$ and this happens 1000 times for a total of one million simulations.

\begin{figure}
\includegraphics[clip, trim={2.5cm 8cm 4cm 7cm}, scale=.5]{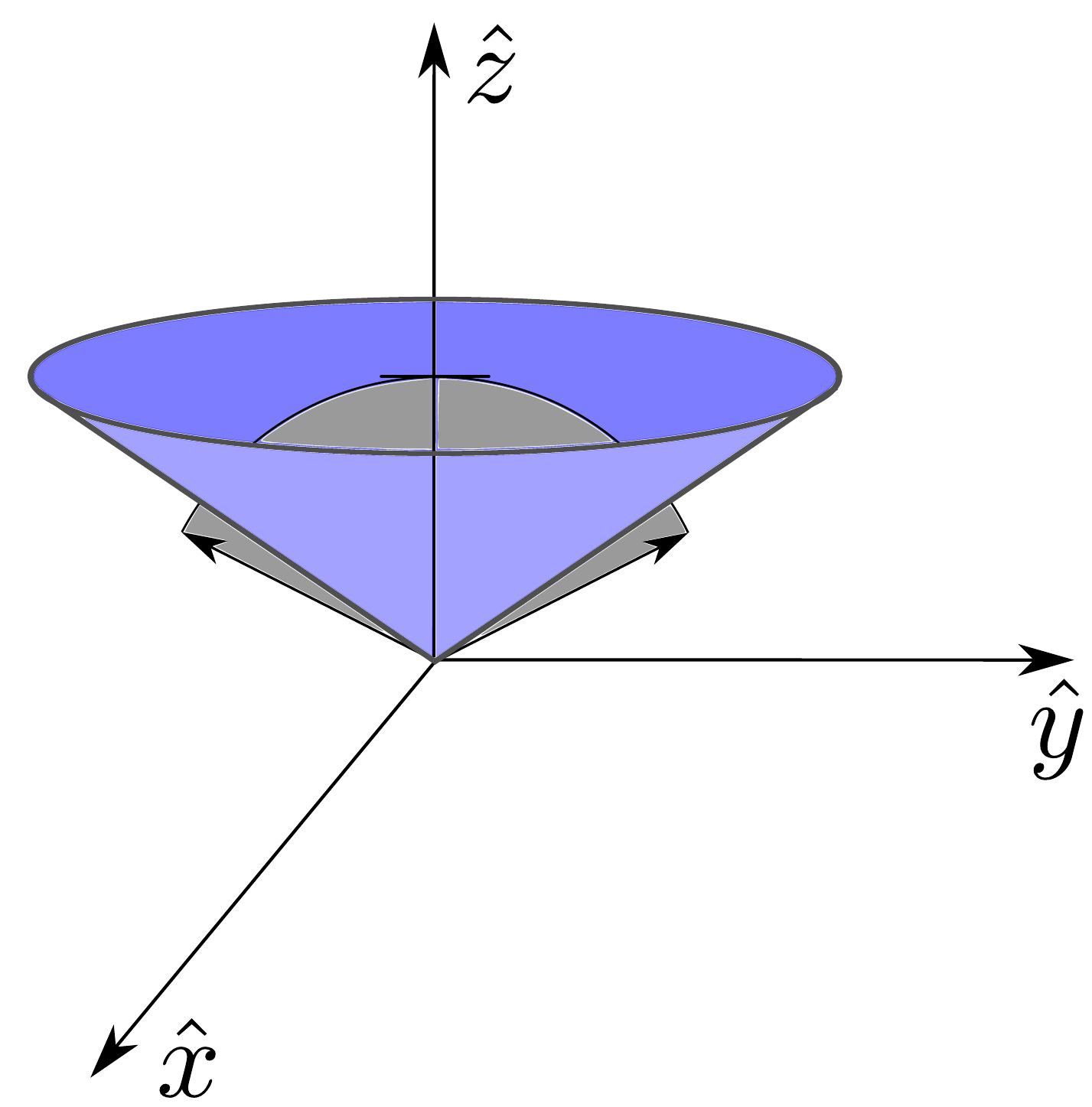}
\caption{\label{fig:Qspin_uncertainty}The standard 3D quantum spin 1/2 result is shown as well as the 2D probability region for the multi-body quantum spin model.}
\end{figure}

The random value of each individual spin component, $\theta_i$, based around $\left\langle \theta_{int} \right\rangle$ is $\pm$1.2 radians.  The visual comparision between this initial uncertainty and the final uncertainty for the quantum model is shown in figure \ref{fig:Qspin_uncertainty} (the angles between the two models are to scale).

As seen in figure \ref{fig:Multi_Body_Model}, the model does not include wrap around boundary conditions.  At first the model included large $N \times N$ quantum spin components, but it was shortly realized that not many sub-elements are needed to obeserve the needed quantum statistics.

\section{Results}
The semi-classical quantum spin interpretation needs to be compared to quantum statistics if the model is to be suggested as a good candidate.  The needed statistical replication, equation \ref{eq:Percentage}, will be used as a first validation (the probability for the spin to evolve into a spin up state will be evaluted).

\begin{figure}
\includegraphics[clip, trim={0cm 0cm 0cm -1cm}, scale=.48]{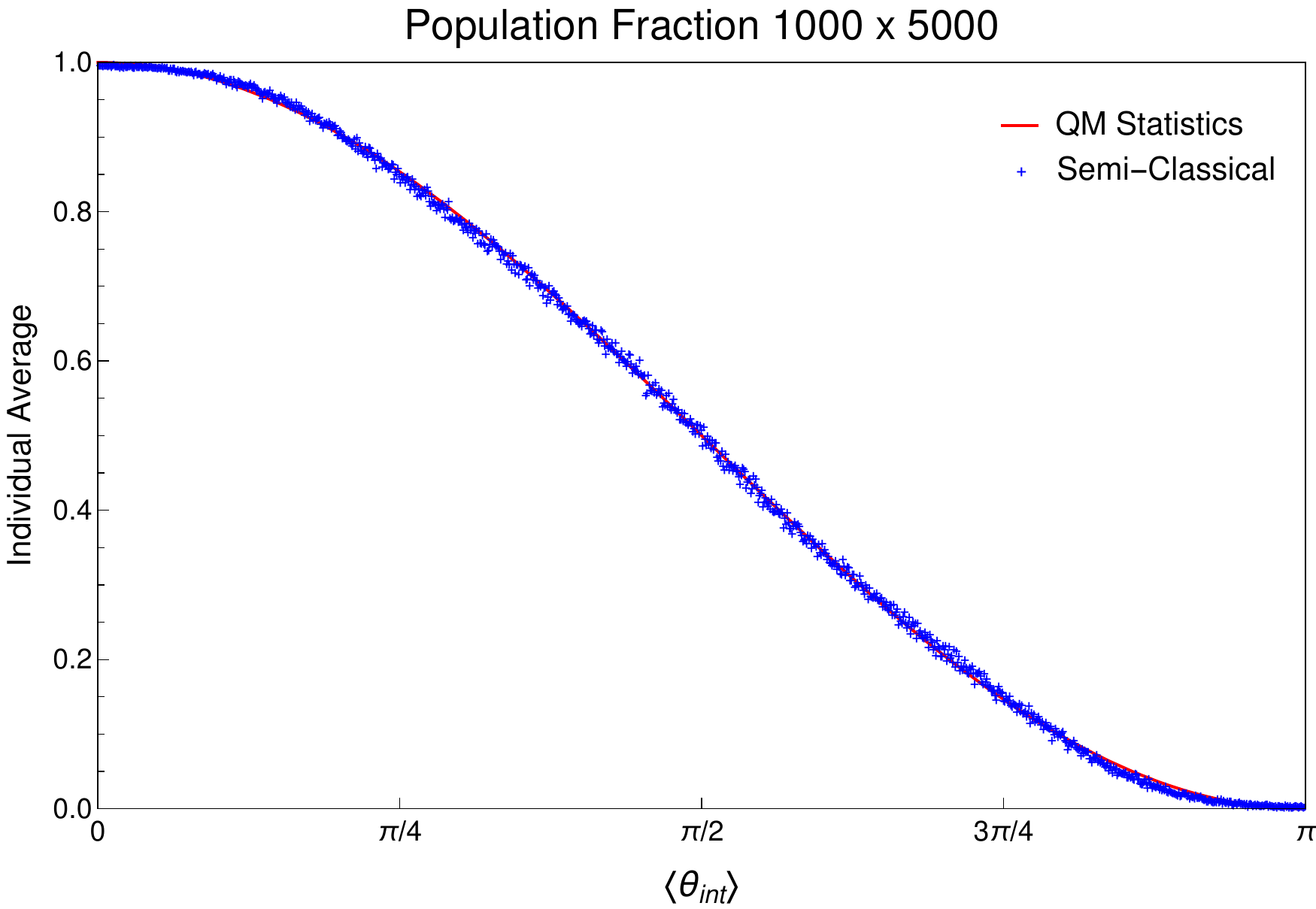}
\caption{\label{fig:QM_comp_anti}Comparision of the well known quantum spin staticstics to the multi-body, anti-ferromagnetic ordering, semi-classical model.  Each blue marker represents a fraction population of 5,000 individual runs.}
\end{figure}

The best quantum mechanical statistical comparision to the multi-body with anti-ferromagnetic ordering model is shown in figure \ref{fig:QM_comp_anti}.  From the figure, the correlation between quantum statistics and the proposed model is strong.

Although a Lyapunov spectrum has yet to be calculated from the Jacobian, a simple divergence separation measurement was taken from two trajectories in phase space differing as $\lvert \delta Z_0 \rvert$:

\begin{equation}
\lvert \delta Z(t) \rvert \approx \textrm{e}^{\lambda t} \lvert \delta Z_0 \rvert.,
\label{eq:lambda}
\end{equation}

\noindent which yielded a positive Lyapunov exponent $\lambda = 1.2$.  This shows that indeed the nonlinear semi-classical system exhibits chaos as required per literature \cite{Thayer_Jafari} (Since the purpose of this paper was to show a semi-classical spin model that replicates quantum statistics, a full analysis of the chaos will be given in a future publication).

\section{Conclusion}

The multi-body, anti-ferromagnetic ordering, semi-classical model as shown in this research shows the ability to replicate quantum mechanical statistics (which up to this point could only be replicated from quantum mechanical wave theory).

Although an energy difference between the spin up and down state is zero in this research, as seen from equation \ref{eq:torque_semi_classical}, the difference in energy levels, verified readily by the Zeeman effect, can be interpreted as being small compared to the coupling of nearest neighbor interactions, $\tau_\textrm{sc} < \tau_\textrm{nn}$.  Therefore to first order, an energy difference as seen in the torques can be ignored.

Furthermore, this alternative representation for a quantum spin can provide a deeper understanding and a more realistic local interpretation for two entangled particles, in a singlet state, commonly known as nonlocal, 'spooky action at a distance'.  Although this research included one quantum spin, when applied to the spin correlation singlet state (as in the EPR/Bell Inequality analysis \cite{Einstein, Bell}) this research predicts that two opposite pairs have statistically opposite outcomes.  Thus, entanglement incorporates quantum statistics, as seen in spin correlated analyses (such as used in Bell inequality studies); however, this result is due to sensitivity to initial conditions.  Consequently, a nonlocal explaination of entanglement is not needed to fully understand spin correlation pairs.

As a result of this analysis and a needed follow-up, the semi-classical quantum spin predicts an entanglement lifetime, or more correctly stated, a correlation-like lifetime, due to chaos.  For example at a short time after creation, particle one will be exposed to something different than particle two.  This will introduce a small shift in trajectory phase space.  From equation \ref{eq:lambda}, the path of a perfectly different outcome will depend on the lyapunov exponent, and thus a lifetime correlation. 

\section{Discussion}

The multi-body model has been shown to replicate quantum statistics and provide a more realistic interpretation of the EPR paradox.  The multi-body model can also help explain quantum energy tunneling without the mediation of virtual photons.  For example, the Rabi oscillation probability of going from a spin up state, 1, to a spin down state, 2, while at resonance, $\omega_0$, can be explained, where

\begin{equation}
P_{1 \rightarrow 2}=\sin^2 \Big( \frac{\omega_1 t}{2} \Big),
\label{eq:prob_rabi}
\end{equation}

\noindent when exposed to a constant dominant magnetic field and a perpendicular oscillating magnetic field perturbation, as:

\begin{equation}
\vec B = B_0 \hat z + B_1(\cos {\omega_0 t} \, \hat x - \sin {\omega_0 t} \, \hat y).
\end{equation}

\noindent Here it should be noted that at some time $t_0$, where $t_0 \ll \pi/2\omega_1$, there exists a non-zero probability that the spin will be found in a down state.  Energy conservation is violated because the system has increased in energy by $2 \mu B_0$, and has previously been resolved by virtual photon mediation.  From the semi-classical model shown, since this energy difference is small compared to nearest neighbor coupling, the conserving energy differece $2 \mu B_0$ comes from nearest neighbor coupling energy.  Thus, energy is conserved without invoking a virtual interaction.

Furthermore, from the uncertainty principle, $\Delta E \, \Delta t \geq \hbar / 2$, a virtual particle is allowed to borrow energy, $\Delta E$, (in the form of virtual particles) for a restricted amount of time, $\Delta t$, \cite{Hey}.  One observable consequence of invoking virtual particle mediation is the quantum spin states that collapse anti-parallel to the magnetic field before one half Rabi cycle, have borrowed energy and their lifetimes should be less than quantum spin states that collapse anti-parallel after one half Rabi cycle (the lifetime being confined by the uncertainty principle).  Although rabi lifetimes should be dependent on the Rabi cycle due to the uncertainty principle, this has not been observed \cite{Norris, Ilichev, Batalov}.

Additionally, the theoretical Rabi oscillation probability, equation \ref{eq:prob_rabi}, is symmetric around one half Rabi cycle and matches closely with experiment \cite{Pla, Koppens, Petta}.  The current interpretation requires virtual particle mediation (for spin states collapsing anti-parallel to the magnetic field) for the first half of a Rabi cycle.  Since the first half of a Rabi oscillation requires virtual particle mediation and the other doesn't, the question naturally arises, ``How can both sides be symmetric, if the process during the collapse is dramatically different?''

The testing of the nonlinear quantum spin model shown in this research will replicate Rabi oscillations.  Although Rabi oscillations are still often misunderstood as purely a quantum effect, since this can be explained as a semi-classical result from linear dispersion and absorption \cite{Zhu}, the state population vector is a classical evolution in time.  Once the evolution has stopped and the system is measured, the same quantum statistics shown in this research are obtained. 

This work was supported by the NASA Goddard Space Flight Center under the cooperative agreement \#NNX13AJ37A, the University of Wyoming Physics and Astronomy department, the Mount Moran HPC cluster at the Advanced Research Computing Center \cite{Mtmoran}, and the Wyoming NASA Space Grant Consortium, NASA Grant \#NNX15AI08H.



\begin{thebibliography}{5}

\bibitem{Thayer_Jafari}Thayer, D.R., and Jafari, F., \textit{Int. J. Ad. Res. Phys. Sc.} \textbf{2}(2) 18-26 (2015)

\bibitem{Thayer}Thayer, D.R., \textit{Int. J. Ad. Res. Phys. Sc.} \textbf{2}(7) 1-18 (2015)

\bibitem{2D_qspin}Heiner, J.J., and Thayer, D.R., \textit{Int. J. Ad. Res. Phys. Sc.} \textbf{4}(3) 4-11 (2017)

\bibitem{New_Zealand}Heiner, J.J., Bodyfelt, J.D., and Thayer, D.R., submitted for publication

\bibitem{Rauch}Rauch, D et al., \textit{Phys. Rev. Lett.} \textbf{121}(8) 080403 (2018)

\bibitem{Bierhorst}Bierhorst, P., et al., \textit{Nature} \textbf{556} 223-226 (2018)

\bibitem{Raffaelli}Raffaelli, F., et al., \textit{Quantum Sci. Technol.} \textbf{3} 025003 (2018)

\bibitem{Einstein}Einstein, A., Podolsky, B., and Rosen, N., \textit{Phys. Rev.} \textbf{47} 777 (1935)

\bibitem{Caves}Caves, C.M., Fuchs, C.A., and Schack, R., \textit{Phys. Rev. A} \textbf{65} 022305 (2002)

\bibitem{Jaynes}Jaynes, E.T., ``Probability in quantum theory'' \textit{Complexity, entropy and the physics of information} 381-403 (1990)

\bibitem{Gallow}Gallow, J.D., ``A Subjectivist's Guide to Deterministic Chance'' \url{https://philarchive.org/archive/GALASG}

\bibitem{Ossiander}Ossiander, M., et al., \textit{Nature} \textbf{561} 374-377 (2018)

\bibitem{Markov}Markov, M.A., Prog. Theor Phys. Suppl., Commemoration Issue for the Thirtieth Anniversary of Meson Theory and Dr. H. Yukawa, p. 85 (1965); \textit{Sov. Phys. JETP} \textbf{24} 584 (1967)

\bibitem{Kadyshevsky}Kadyshevsky, V.G., \textit{Nucl. Phys.} \textbf{B141} 477 (1978)

\bibitem{Mateev}Kadyshevsky, V.G. and Mateev, M.D., \textit{Phys. Lett.} \textbf{B106} 139 (1981)

\bibitem{Rodionov}Rodionov, V.N. \textit{Physica Scripta} \textbf{90}(4) 2015

\bibitem{Hooft}Hooft, G, ``The Evolution of Quantum Field Theory'' \textit{The Standard Theory of Particle Physics} Advanced Series on Directions in High Energy Physics \textbf{26} 1-27 (2015)

\bibitem{Bell}Bell, J.S., \textit{Physics} \textbf{1} 195 (1964)

\bibitem{Hey}Hey, T., and Walters, P., ``The New Quantum Universe'' Cambridge University Press (2004)

\bibitem{Norris}Norris, T.B., et al., \textit{Phys. Rev. B} \textbf{50}(19) (1994)

\bibitem{Ilichev}Il'ichev, E., et al., \textit{Phys. Rev. Lett.} \textbf{91}, 097906 (2003)

\bibitem{Batalov}Batalov, A., et al., \textit{Phys. Rev. Lett.} \textbf{100}, 077401 (2008)

\bibitem{Pla}Pla, J.J., et al., \textit{Nature} \textbf{489}, 541-545 (2012)

\bibitem{Koppens}Koppens, F.H.L., et al., \textit{Nature} \textbf{442}, 766-771 (2006)

\bibitem{Petta}Petta, J.R., et al., \textit{Science} \textbf{309}(5744), 2180-2184 (2005)

\bibitem{Zhu}Zhu, Y., Gauthier, D.J., Morin, S.E., Wu, Q., Carmichael, H.J., and Mossberg, T.W., \textit{Phys. Rev. Lett.} \textbf{64}(21) (1990)

\bibitem{Mtmoran}Advanced Research Computing Center. 2012. Mount Moran: IBM System X cluster. Laramie, Wy: University of Wyoming. \url{http://n2t.net/ark:/85786/m4159c}

\end{thebibliography}
\end{document}